\documentclass[proceedings]{JHEP}

\usepackage{amsmath,euscript,array,amssymb,cite} 

\def\aD{{\dot\alpha}}
\def\bD{{\dot\beta}}
\def\M{{{\cal M}'}}
\def\N{{\cal N}}

\def\sst{\scriptscriptstyle}
\def\det{{\rm det}}

\def\D{{\cal D}}

\newcommand{\BN}{\boldsymbol{N}}
\newcommand{\Bk}{\boldsymbol{k}}
\newcommand{\BL}{\boldsymbol{L}}

\newcommand{\BX}{\boldsymbol{X}}
\newcommand{\Balpha}{{\boldsymbol{\alpha}}}
\newcommand{\Bsigma}{{\boldsymbol{\sigma}}}
\newcommand{\Bvarphi}{{\boldsymbol{\varphi}}}

\def\Dbarslash{\,\,{\raise.15ex\hbox{/}\mkern-12mu {\bar\D}}}
\def\Dslash{\,\,{\raise.15ex\hbox{/}\mkern-12mu \D}}
\def\delslash{\,\,{\raise.15ex\hbox{/}\mkern-9mu \partial}}
\def\delbarslash{\,\,{\raise.15ex\hbox{/}\mkern-9mu {\bar\partial}}}

\def\ms{{\mathfrak M}}
\def\Z{{\EuScript Z}}

\def\VEV#1{\Big\langle #1\Big\rangle}
\def\Vev#1{\big\langle{#1}\big\rangle}

\newcommand{\EQ}[1]{\begin{equation} #1 \end{equation}}

\newcommand{\SP}[1]{\begin{equation}\begin{split} #1 \end{split}\end{equation}}

\conference{Non-perturbative Quantum Effects 2000}

\title{A Brief History of the Stringy Instanton}

\author{Nick~Dorey$^{a}$, Timothy J.~Hollowood$^{a}$
and Valentin V.~Khoze$^b$\\
$^a$Department of Physics, University of Wales Swansea,
Swansea, SA2 8PP, UK\\
$^b$Department of Physics and IPPP, University of Durham,
Durham, DH1 3LE, UK\\
E-mail: {\tt n.dorey@swan.ac.uk}, {\tt t.hollowood@swan.ac.uk},
{\tt valya.khoze@durham.ac.uk}}

\abstract{The arcane ADHM construction of Yang-Mills instantons can be
very naturally understood in the framework of D-brane dynamics in
string theory. In this point-of-view, the mysterious auxiliary
symmetry of the ADHM construction arises as a gauge symmetry and the
instantons are modified at short distances where string effects become
important. By decoupling the stringy effects, one can recover all the
instanton formalism, including the all-important 
volume form on the instanton moduli
space. We describe applications of the instanton calculus to the
AdS/CFT correspondence and higher derivative terms in the D3-brane
effective action. In these applications, we are starting to 
uncover an interesting
relation between instanton partition functions, the Euler
characteristic of instanton moduli space and modular symmetry.
We also describe how it is now possible to do multi-instanton
calculations in gauge theory and we resolve an old puzzle involving the
gluino condensate in supersymmetric QCD.}


\preprint{{\tt hep-th/0010015}}

\begin{document}

\section{Introduction}\label{sec:S1}

There has been a wealth of progress in understanding semi-classical
effects in supersymmetric gauge theories in the last few years. In
this review we shall be concentrating on instantons and one of
the main goals is show how naturally Yang-Mills instantons appear in
string theory.

Instantons are solutions of the classical equations of gauge theory
with finite action. Charge $k$ instanton effects come along with a
factor
\SP{
&\exp\big(-\tfrac1
{2g^2}\int F\wedge ^*F+\tfrac{i\vartheta}{16\pi^2}\int F\wedge
F\big)\\
&=\exp \big(-\tfrac{8\pi^2 k}{g^2}+ik\vartheta\big)\equiv
e^{2\pi ik\tau}\equiv q^k\
\thicksim \Lambda^{kb_1}\ ,
\label{insta}
}
where $b_1$ is the first coefficient of the beta function.
Certain quantities in SUSY gauge theory are protected
by powerful non-renormalization theorems or
``holomorphy''. In these cases, instanton contributions are exact in
the sense that there are no perturbative corrections since these would
involve a series in the non-holomorphic quantity $g^2$.
The classic example of such a protected quantity is the gluino
condensate in $\N=1$ SUSY gauge theory: for gauge group $SU(N)$ an
instanton calculation gives \cite{scgc}
\EQ{
\VEV{\frac{\lambda^2}{16\pi^2}}=\frac{2\Lambda^3}
{\big[(N-1)!(3N-1)\big]^{1/N}}\ .
\label{propex}
}
Actually this appears to be a charge $k=\tfrac1N$ effect based on the power
of $\Lambda$, so what is calculated is the instanton contribution to
the $N$-point function which is independent of the insertion points by
a SUSY Ward identity. Clustering is then invoked (taking
into account averaging over the $N$ physically equivalent vacua of
the theory) to extract \eqref{propex}.

Unfortunately, we have recently learned that this procedure does not
give the correct answer for the gluino condensate
\cite{HKLM}.\footnote{One of the reasons for believing in the result
is due to a topological field theory style argument 
along the lines that the semi-classical
approximation should be exact. However, this only applies to the
theory on a compact spacetime, like the torus and {\it not\/} on
${\mathbb R}^4$.} 
Suspicion should have been aroused from the start, after all
an instanton
calculation is a semi-classical method, whereas the theory in question
is in a strongly-coupled confining phase. 
So although there are no perturbative
corrections around the instanton contribution there can be other
non-perturbative, but non-instanton, contributions. How can we be sure
that other configurations contribute to the gluino condensate? We know
this because in \cite{HKLM} we
calculated for the first time multi-instanton contributions to
multi-point functions of the gluino
and showed that clustering is violated: a sure sign that other
configurations must contribute. Specifically, we calculated the
contribution of $k$ instantons to the $kN$-point function, at large
$N$ (which is enough to prove the point), and showed in this limit
\EQ{
\VEV{\frac{\lambda^2(x_1)}{16\pi^2}\times\cdots\times
\frac{\lambda^2(x_{kN})}{16\pi^2}}^{\tfrac1{kN}}\propto k\ ,
}
instead of being independent of $k$, if clustering was respected in
the instanton sector.

The message of this work is two-fold: firstly we have demonstrated that
multi-instanton calculations are now technically feasible,
particularly in the large $N$ limit where a number of important
simplifications occur. Secondly, instanton calculations will be valid
only in a weakly-coupled phase. This last point looks rather
restrictive; however, this is not the case and one can infer the value
of a quantities like the gluino condensate in a strongly coupled phase
using a two-stage procedure.

The idea is to modify the theory so that there is a new coupling
constant which in some limit drives the theory into weak coupling, but
which in the opposite limit returns one smoothly to the confining
phase. Under suitable circumstances the gluino condensate will be
holomorphic in this new coupling and the weak-coupled result can be
analytically continued to strong coupling. There are at least two ways
to achieve this and both yield the correct answer for the gluino
condensate
\EQ{
\VEV{\frac{\lambda^2}{16\pi^2}}=\Lambda^{3}\ .
}

{\bf Method 1.} \cite{wcgc}
Add matter fields to completely break the gauge
group. For example, we can add $N-1$ chiral multiplets in the
$\BN+\bar{\BN}$. The resulting theory is in a weakly-coupled Higgs phase
for small values of the masses. When the masses go to infinity, the
matter fields decouple and the theory is continuously connected to the
confining phase of the pure gauge theory. The gluino condensate can
now be calculated reliably. In this case the 1-point function is a
one-instanton effect.

{\bf Method 2.} \cite{cyl}
Put the theory\footnote{This method works for and  arbitrary gauge
group $G$, with Lie algebra ${\mathfrak g}$,
and we will take $r$ to be the rank.}
on a cylinder ${\mathbb R}^3\times
S^1$.
In this case, the gauge field can have a Wilson line around the
$S^1$. This breaks the gauge group to $U(1)^r$ and so, generically, the theory
is in a Coulomb phase. For small value of the radius the theory is
weakly coupled and the gluino condensate can be calculated
reliably. The result can then be continued to large $R$ and gives the
value of the gluino condensate in the uncompactified theory.
Interestingly, on the cylinder the topological charge is
not constrained to be integer and there are other finite action
configurations which arise as monopoles in the gauge theory whose
world-lines wrap the $S^1$. The gluino condensate now receives
contributions from monopoles. For a general gauge group there are
$r$ fundamental monopoles (those which are not
composite configurations) whose magnetic charges
are proportional to the co-simple roots $\Balpha^*_i$ of ${\mathfrak
g}$. However, on the
cylinder there is an additional solution, the ``affine'' monopole,
whose magnetic charge is the lowest co-root $\Balpha_0^*$
\cite{Lee:1997vp}. This
solution is special to the cylinder
since it depends non-trivially on the coordinate
around the circle. Amazingly, an instanton in the theory on the
cylinder is a composite
configuration consisting of one of
each of the $r+1$ fundamental monopoles. For small
radius the theory is in a weakly-coupled Coulomb phase and the
fundamental monopole contribute to a superpotential in the low energy
effect action of the 3-dimensional $U(1)^r$ gauge theory. This
superpotential depends on an $r$-vector superfield $\BX$
whose scalar component is
\EQ{
\Bsigma+\tau\Bvarphi\ ,
}
where $\Bsigma$ is the dual $U(1)^r$ gauge field and $\Bvarphi$ is the
Wilson line. The superpotential has the
form of an affine Toda potential for $({\mathfrak g}^{(1)})^*$,
matching calculations via M theory \cite{Katz:1997th}:
\EQ{
W(\BX)\thicksim \sum_{j=1}^r\tfrac2{\Balpha_j^2}e^{i\Balpha_j^*\cdot\BX}
+q\tfrac2{\Balpha_0^2}e^{i\Balpha_0^*\cdot\BX}\ .
}
where $\Balpha^*=2\Balpha/\Balpha^2$ are the dual roots.
This potential has $c_2$ SUSY vacua (in agreement with the Witten
index) and the magnitude of the gluino condensate in each of the vacua is
\EQ{
\VEV{\frac{\lambda^2}{16\pi^2}}=\frac{\Lambda^3}{\prod_{j=0}^r(k_j^*
\Balpha^2_j/2)^{k_j^*/c_2}}\ ,
}
where $k_j^*$ are the dual Kac labels.
Interestingly in each vacua the monopole carry a fraction $1/c_2$ of
topological charge which means that they realize the old idea that an
instanton is made up of constituents.

The connection of superpotentials in SUSY gauge theories
compactified on a cylinder and integrable theories is more
general. The most remarkable example is the mass deformed $\N=4$
theory (or the $\N=1^*$ theory) \cite{Dorey:1999sj}.
The $\N=4$ theory consists of an $\N=1$
vector multiplet and 3 adjoint-valued $\N=1$ chiral multiplets, each of
which can be given a mass $m_{1,2,3}$ which breaks $\N=4\to1$.
For gauge group $SU(N)$, the superpotential for the theory
compactified on the cylinder is the complexified potential of the
$N$-body elliptic Calogero-Moser system:
\EQ{
W(\BX)\thicksim m_1m_2m_3\sum_{\Balpha}\wp(\Balpha\cdot\BX)\ .
}
When expanded in $q$, terms can be identified
with particular configurations involving monopoles and
instantons \cite{Dorey:1999sj}. This superpotential has some
remarkable modular properties and one can extract a wealth of
information from it \cite{mdn4}.

\section{Multi-Instantons and ADHM}\label{sec:S2}

We now need to get to grips with the calculus of multi-instantons on
${\mathbb R}^4$ as described by ADHM \cite{ADHM}. On first
exposure, the ADHM construction looks rather {\it ad-hoc\/}; however
in the following section we shall describe 2 ways to interpret it.

We start by considering a single instanton in $SU(N)$.
This is constructed by taking an $SU(2)$ instanton, which has
a scale size and position in ${\mathbb R}^4$, and then orientating it
inside $SU(N)$, which involves $4N-5$ additional ``coset''
parameters. Rather perversely, we want to describe these moduli in the
following way: firstly $a'_n$, which is (minus) the position of the
instanton. To this we add 2 complex $N$ vectors $w_{u\aD}$ subject
to the 3 constraints
\EQ{
(\tau^c)^\aD_{\ \bD}\bar w^\bD w_\aD=0\ .
\label{oic}
}
(Obviously, we could easily solve the constraints in this case.)
The instanton solution is actually
independent of an auxiliary $U(1)$ which rotates
$w_\aD$ by a phase. The physical meaning of the parameters is
\EQ{
\begin{matrix}a'_n &\longrightarrow& -\text{position in }{\mathbb R}^4\\
\rho^2=\bar w^\aD w_\aD &\longrightarrow& \text{size}^2\\
\rho^{-2}w_{u\aD}(\tau^c)^\aD_{\ \bD}\bar w^\bD_v
&\longrightarrow&
SU(2)\subset SU(N)\end{matrix}
}

Multi-instantons are described by a non-abelian generalization of this
construction. The instanton position $a'_n$ becomes a 4-vector of $k\times k$
hermitian matrices and there are $2k$ $N$-vectors $w_{ui\aD}$,
$i=1,\ldots,k$. The
generalization of \eqref{oic} is the famous set of ADHM constraints:
\EQ{
{\EuScript B}^c\equiv(\tau^c)^\aD_{\ \bD}\big(\bar w^\bD w_\aD+\bar
a^{\prime\bD\alpha}a'_{\alpha\aD}\big)=0\ ,
\label{ADHM}
}
where $a'_{\alpha\aD}=a'_n\sigma^n_{\alpha\aD}$. The moduli space of
$k$ instantons, $\ms_{k,N}$, is then given by
$\{a'_n,w_\aD\}$ modulo the ADHM constraints
and modulo an auxiliary $U(k)$ symmetry which acts as $w_\aD\to w_\aD
U$, $a'_n\to U^\dagger a'_nU$.

Before we leave this section, let us consider three important
things.

(i) In a supersymmetric theory, instantons also have
Grassmann moduli which arise from the fermion fields (see
\cite{wcgc,CFGT,KMS,MO3}). In a SUSY gauge theory with $\N$
supersymmetries ($\N=1,2,4$), there are $\N$ gluino fields and
the corresponding Grassmann collective coordinates are  $k\times k$
matrices $\M^A_\alpha$, $k\times N$ matrices $\mu^A$ and
$N\times k$ matrices $\bar\mu^A$, where $A=1,\ldots,\N$. These are
subject to fermionic analogues of the ADHM constraints:
\EQ{
{\EuScript F}^A_\aD\equiv \bar\mu^Aw_\aD+\bar w_\aD\mu^A+
[\M^\alpha,a'_{\alpha\aD}]=0\ .
}

(ii)
In order to do instanton calculations, we need to known how
to change variables in the path integral from the fields to the
collective coordinates. A direct approach to this problem has only
been achieved in the cases $k=1,2$ \cite{OSB,MO12}. An alternative
and tractable approach \cite{meas12} is
to use the symmetries of the theory, and in this respect SUSY is a very
powerful symmetry, along with cluster decomposition to deduce the
measure on the SUSY instanton moduli space at arbitrary $k$. The
resulting expression for the measure is fortunately rather simple:
\EQ{
\Z_{k,N}=\int \frac{d^4a'\,d^2w\,d^{2\N}\M\,d^\N\mu\,d^\N\bar\mu
\,\delta({\EuScript B}^c)\,\delta({\EuScript F}^A_\aD)}
{{\rm Vol}\,U(k)\big(\det\BL
\big)^{\N-1}}\ .
\label{meas}
}
Here $\BL$ is an operator on $k\times k$ matrices:
\EQ{
\BL\cdot\Omega=\{\bar w^\aD w_\aD,\Omega\}+[a'_n,[a'_n,\Omega]]\ .
}
Given the measure in any of the supersymmetric theories, the measure
in QCD ($\N=0$) can be deduced by giving masses to the fermions
and then using renormalization group decoupling.
Remarkably the resulting
expression is also given by the formula \eqref{meas}. It is
now possible to write down the $k$ instanton measure in pure QCD,
including the one-loop fluctuation determinants, since these
can be extracted from the old instanton literature \cite{Jack:1980rn}.
Since, to our knowledge this has never been written down,
we do it here
\EQ{
\Z_{\rm QCD}=\int \frac{d^4a'\,d^2w
\,\delta({\EuScript B}^c)}
{{\rm Vol}\,U(k)}\cdot{\rm det}\BL\cdot(\det\Delta)^{-2}
\ ,
\label{mqcd}
}
where the fluctuation determinant is
\EQ{
(\det\Delta)^{-2}=\mu^{-kN/3}(\det\BL)^{-1}e^{-\tfrac1{24\pi^2}
\int d^4x(I_1+I_2+I_3)}
\ ,
}
where $I_c(x)$ are all functions of the $k\times k$
matrix
\EQ{
f^{-1}(x)=\tfrac12\bar w^\aD w_\aD
+(a'_n+x_n1_{\sst [k]\times[k]})(a'_n+x_n1_{\sst
[k]\times[k]})\ ,
}
given by
\SP{
I_1(x)&={\rm tr}_k\big(f\partial_nf^{-1}  f\partial_nf^{-1}
f\partial_mf^{-1}  f\partial_mf^{-1}\\&\qquad\qquad-20f^2\big)
+\frac{4k}{(1+x^2)^2}\ ,\\
I_2(x)&=\int_0^1 dt\,
\epsilon_{mnpq}{\rm tr}_k\big(\tilde f\partial_t \tilde f^{-1}
\tilde f\partial_m \tilde f^{-1} \\&\qquad\qquad\times
 \tilde f\partial_n \tilde f^{-1}
\tilde f\partial_p \tilde f^{-1}  \tilde f\partial_q \tilde
f^{-1}\big)\ ,\\
I_3(x)&=\log\det\,
f\ \square^2\log\det\, f\ .
}
Here, $t$ is an auxiliary variable and $\tilde f(x,t)$ is the $k\times k$
dimensional matrix derived from $f(x)$:
\EQ{
\tilde f^{-1}(x,t)=tf^{-1}(x)+(1-t)(1+x^2)1_{\sst [k]\times[k]}\ .
}
In \eqref{mqcd}, $\mu$ is the mass parameter of the Pauli-Villars
regularization scheme.
Of course we should emphasize that QCD is {\it not\/} weakly coupled and
in the light of our previous discussion, we should be rather careful
in using the instanton approximation in this context.

In an $\N=4$ SUSY gauge theory the measure \eqref{meas} is not the
complete story because in these theories the action evaluated on the
instanton solution is not just the constant \eqref{insta}. In
these theories all but the 8 SUSY and 8 superconformal fermion zero
modes, which are
protected  by the corresponding symmetries, are lifted beyond linear
order at the classical level
by the Yukawa interactions of the theory \cite{MO3}. This leads to
a 4-fermion term in the instanton action:
\SP{
\tfrac{\pi^2}{2g^2}
\epsilon_{ABCD}&{\rm tr}_k(\bar\mu^A\mu^B+\M^{\alpha A}\M^B_\alpha)\\
&\times\BL^{-1}
(\bar\mu^C\mu^D+\M^{\beta C}\M_\beta^D)\ .
\label{4fc}
}

(iii) The moduli space $\ms_{k,N}$ is not a
smooth manifold: it has orbifold-type singularities that occur when
$U(k)$ does not act freely. Physically these are points
where an instanton shrinks to zero size, {\it i.e.\/}~$w_{iu\aD}=0$
for a given $i$. We can illustrate this for the case of a single
instanton in $SU(2)$. In this case,
$\ms_{1,2}={\mathbb R}^4\times{\mathbb R}^4/{\mathbb Z}_2$, where
${\mathbb R}^4$ corresponds to position of the instanton
while the angular coordinates of the second ${\mathbb R}^4$
parameterize the $SU(2)$ gauge orientation and finally the radial coordinate
of this factor is the scale size.
It is important to emphasize that these singularities are
{\em not\/} evidence of any sickness in the instanton calculus. In
fact when calculating the instanton contribution to any physical
quantity in field theory these short-distance singularities are prefectly harmless.

There is a natural way to smooth, or blow up, the singularities of
$\ms_{k,N}\to\ms_{k,N}^{(\zeta)}$:
simply modify the ADHM constraints by adding a term proportional to
the identity matrix to the right-hand side:
\EQ{
{\EuScript B}^c
\equiv(\tau^c)^\aD_{\ \bD}\big(\bar w^\bD w_\aD+\bar
a^{\prime\bD\alpha}a'_{\alpha\aD}\big)=\zeta^c1_{\sst[k]\times[k]}\ .
\label{mADHM}
}
The new term prevents any component $w_{ui\aD}\to0$ and so instantons
cannot shrink to zero size. For example, in the case $k=1$ and
$N=2$ described above, it is possible to show that the orbifold factor
${\mathbb R}^4/{\mathbb Z}_2$ becomes the Eguchi-Hanson manifold.
Remarkably, the smoothed moduli space $\ms_{k,N}^{(\zeta)}$
describes instantons in non-commutative gauge theory on a  spacetime
with non-commuting coordinates \cite{Nekrasov:1998ss}:
\EQ{
[x_n,x_m]=-i\bar\eta^c_{nm}\zeta^c\ ,
}
where $\bar\eta^c_{nm}$ is a 't~Hooft eta symbol.

\section{Meaning of ADHM}\label{sec:S3}

This all seems rather mysterious: we have a curious set of data and an
auxiliary $U(k)$ symmetry. The first point to make is that the ADHM
constraints are generally intractable: one simply cannot find a
solution for $k>3$. However, recently we found a way to solve the constraints
for arbitrary $k$ when $N\geq 2k$ as we describe later.
The rather arcane ADHM construction can now be understood in two
apparently different, althought intimately related, ways:

{\bf The Math Way.}
\cite{Hitchin:1987ea} The ADHM construction is an example of a
hyper-K\"ahler quotient. One starts with ${\mathbb R}^{4kN+4k^2}$, which
is naturally hyper-K\"ahler with a flat metric, and then imposes a
triplet of constraints of the type \eqref{ADHM}. Finally one mods out by
an auxiliary symmetry, in this case $U(k)$,
leaving a hyper-K\"ahler manifold of dimension
$4kN$. It turns out that the measure on the ADHM moduli space
(\eqref{meas} with $\N=0$)
follows geometrically from this quotient construction
as the measure induced on $\ms_{k,N}$ by
the flat measure on ${\mathbb R}^{4kN+4k^2}$. Finally,
the smoothed space described by the modified ADHM constraints
\eqref{mADHM} is a natural deformation of the quotient construction
which preserves the hyper-K\"ahlarity.

{\bf The Physics Way.} \cite{MO3,DW}
The ADHM construction naturally arises in the
dynamics of D-branes in string theory. The low energy collective
dynamics of $N$ coincident D$(p+4)$-branes in Type II string theory
is described by a $U(N)$ SUSY gauge theory in $p+5$-dimensions with 16
supercharges. An instanton in the world-volume
theory of the D$(p+4)$-branes is a soliton which
has 4 transverse directions in the higher dimensional brane, {\it
i.e.\/}~is some kind of $p$-brane.
The remarkable thing is that it is precisely a D$p$-brane bound to the
D$(p+4)$-brane. In general $k$ D$p$-branes
bound to the $N$ higher dimensional D$(p+4)$-branes correspond to a charge
$k$ instanton in a $U(N)$ SUSY gauge theory.

In order to see how this plays out,
we have to consider the low energy collective
dynamics of the D$p$-branes. This is described by a SUSY $U(k)$ gauge
theory with 16 supercharges, but with additional matter fields
arising from the higher dimensional branes which break half of these
supersymmetries. To be more specific, let
us suppose that $p=3$. In this case a theory with 16 supercharges is
$\N=4$ SUSY gauge theory. Let us analyse the spectrum of fields in
terms of $\N=1$ supermultiplets. Along with the $\N=1$ vector
multiplet containing the gauge field, there are 3 adjoint-valued chiral
superfields $\Phi$, $X$ and $\tilde X$. The 6 real scalars of these
chiral multiplets describe the transverse positions of the
D3-branes and in particular $X$ and $\tilde X$ describe the positions
of the D3-branes within the D7-branes, while $\Phi$ describes the
separation between the D3- and D7-branes.
Open string going between the D3-branes and D7-branes
give rise a $N$ chiral multiplets $Q$ and $\tilde Q$ in, respectively,
the $\Bk$ and $\bar{\Bk}$
representations of the gauge group. The resulting theory has $\N=2$
supersymmetry and $X$ and $\tilde X$ form an adjoint hypermultiplet while
$Q$ and $\tilde Q$ form $N$ fundamental hypermultiplets.

This gauge theory then describes the low energy dynamics of the
D3-branes (in the presence of D7-branes).
Let us consider the space of vacua of this theory.
The theory has a Higgs branch where the gauge group is
completely broken (the scalar
components of) $\Phi=0$ and $Q$, $\tilde Q$, $X$ and $\tilde X$ are
non zero. The equations describing the Higgs branch follow from the $D$
and $F$-flatness conditions and these precisely the ADHM constraints
\eqref{ADHM} with the identifications
\EQ{
w_\aD=\begin{pmatrix}
Q^\dagger\\ \tilde Q\end{pmatrix}\ ,\qquad
a'_{\alpha\aD}=\begin{pmatrix} X^\dagger& \tilde X\\
-\tilde X^\dagger & X \end{pmatrix}\ .
}
Hence there is a natural identification of $\ms_{k,N}$ and the
Higgs branch of our $\N=2$ gauge theory. Notice that this gauge
theory, with gauge group $U(k)$,
is {\em not\/} the original $\N=4$ gauge theory that lives on
the D7-branes, which has gauge group $U(N)$.
The Higgs branch describes a situation in which the D3-branes lie inside
the D7-branes ($\Phi=0$). On the contrary the Coulomb branch, on which
$Q=\tilde Q=0$, while $\Phi$, $X$ and $\tilde X$ are non zero,
describes a situation in which the D3-branes have moved off the
D7-branes. There are mixed branches which describe situations in which
some of the D3-branes are on the D7-branes while some have moved off
into the bulk. The points where $Q_i$ and $\tilde Q_i$ go to zero
connect the different phases and
correspond to points where the D3-branes can move off into the
D7-branes. These are
precisely the points where an instanton shrinks to zero size.
So in a certain respect, that we will make explicit shortly, this stringy
context leads to a certain resolution of the orbifold singularities of
$\ms_{k,N}$.

However, there is more to this than an identification between
$\ms_{k,N}$  and the Higgs branch of the gauge theory. If we
dimensionally reduce the system to $p=-1$, so that we are describing
a system of D-instantons and D3-branes, then the the partition
function of the $U(k)$ gauge theory (which is now a $0$-dimensional
field---or matrix---theory) can be identified with the measure on the
ADHM moduli space in the limit where bulk effects decouple from the
branes, $\alpha'\to0$. The 4D gauge
field and $\Phi$ can be amalgamated
into $\chi_a$, $a=1,\ldots,6$, an adjoint-valued
6-vector. The bosonic part of the partition function is
\SP{
&\Z_{k,N}=
\int \frac{d^4a'\,d^2w\,d^6\chi\,d^3D}{{\rm Vol}\,U(k)}
\,\exp\big(-{\rm
tr}_k\chi_a\BL\chi_a\\
&+\alpha^{\prime4}{\rm tr}_k[\chi_a,\chi_b]^2
-2\alpha^{\prime4}{\rm
tr}_kD^2+i{\rm tr}_kD^c{\EuScript B}^c\big)\times\cdots\ .
\label{meas2}
}
Here, $D^c$ is an adjoint-valued 3-vector that arises as an auxiliary
field of the 4D theory. Now if we
take $\alpha'=0$, then the integral over $\chi_a$ is Gaussian and
gives rise to a factor $(\det\BL)^{-3}$, while the $D^c$
are nothing but Lagrange multipliers for the ADHM constraints! Notice that
the resulting partition function in this limit gives precisely the the
bosonic parts of the measure on the ADHM moduli space in an $\N=4$
SUSY theory \eqref{meas}.

If we don't take the $\alpha'=0$ limit, then in a sense we resolve the
singularities of $\ms_{k,N}$
since the $D^c$ no longer act as Lagrange
multipliers for the ADHM constraints, rather, the constraints are
smeared over a scale $\sqrt{\alpha'}$.
How does this kind of resolution relate to the blow-up
$\ms_{k,N}^{(\zeta)}$? The
modifications of the ADHM constraints by the parameters $\zeta^c$ can naturally
be incorporated into the stringy construction since they correspond to
Fayet-Illiopolos (FI) couplings in the $U(k)$ gauge theory, {\it
i.e.\/}~add $-i\zeta^c{\rm tr}_kD^c$ to the exponent in \eqref{meas2}.
There are
consequently two different ways to smooth $\ms_{k,N}$, via stringy
corrections or via FI couplings; however, we shall argue later that
they lead to the same effect.

Before we leave this section there are three further issues that we
mention.

(i) It is important that $\chi_a$ also couples to a fermion bilinear:
\EQ{
\Sigma^a_{AB}{\rm tr}_k\chi_a(\bar\mu^A\mu^B+\M^{\alpha A}\M_\alpha^B)\ ,
}
for, when $\alpha'=0$ and $\chi_a$ is integrated-out, a
the 4-fermion interaction \eqref{4fc} is generated.

(ii) Hitherto, we have been considering the situation where
the $N$ D$(p+4)$-branes are coincident; however, what happens when they
separate? Consider the case with $p=-1$. From the point-of-view of
the D3-branes the answer is straightforward: the scalars which
correspond to the positions of the branes gain a VEV 
$\langle\varphi_a\rangle$, a 6-vector of
$N\times N$ matrices, and one moves out onto the Coulomb
branch of the $U(N)$ gauge theory. This effect is then easily
incorporated into the D-instanton $U(k)$ theory, by modifying the
following couplings:
\EQ{
w_\aD\chi_a\to w_\aD\chi_a+\langle\varphi_a\rangle w_\aD\ ,\
\mu^A\chi_a\to \mu^A\chi_a+\langle\varphi_a\rangle\mu^A\ .
\label{vev}
}
These couplings have the form of mass terms for $w_\aD$.
It turns out that the new couplings precisely reproduce the
{\it constrained instanton formalism\/}
\cite{Affleck:1983rr} that describes instantons in theories with VEVs.
The effect of the extra coupling to the VEVs is to suppress
instantons of
large size in the instanton measure and superconformal invariance
is explicitly broken.

(iii)
It is worth commenting on the case when $N=1$ and the
original gauge theory has gauge
group $U(1)$. It is well known that abelian theories do not have
instantons; however, we can still define the ADHM construction.
In this case, the ADHM constraints are explicitly solved by taking
$w_\aD=0$ and $a'_n=-{\rm diag}(X^1_n,\ldots,X^k_n)$. In other words,
these ``abelian instantons'' are point like and moreover
\EQ{
\ms_{k,1}={\rm Sym}^k\big({\mathbb R}^4\big)\ .
}
This space has singularities whenever 2 instantons coincide.
However, one finds that the gauge potential that arises from the ADHM
data is pure gauge. Nevertheless, when we modify the ADHM construction
as in \eqref{mADHM}, in other words consider instantons in a
non-commutative theory, then the abelian instanton solutions become
non-trivial. The deformed space $\ms_{k,1}^{(\zeta)}$ is smooth; for
example, $\ms_{2,1}={\mathbb R}^4\times{\mathbb R}^4/{\mathbb Z}_2$,
while the deformation replaces the orbifold factor with the
Eguchi-Hanson manifold. So we see $\ms_{1,2}=\ms_{2,1}$ and
$\ms_{1,2}^{(\zeta)}=\ms_{2,1}^{(\zeta)}$, a property that does {\it
not\/} generalize to $N>1$ and $k>2$.

\section{Calculations with Instantons}\label{sec:S4}

In this section we shall summarize a number of interesting
applications of the instanton calculus. We will primarily be interested in
the $\N=4$ theory with possible stringy corrections, FI couplings
and VEVs.

\subsection{The AdS/CFT correspondence}\label{sec:S5}

The AdS/CFT correspondence realizes the old idea that string theory
describes large-$N$ of gauge theory \cite{Aharony:2000ti}.
In fact it is much stronger:
$\N=4$ SUSY gauge theory is equivalent to Type IIB string theory
compactified on $AdS_5\times S^5$. As usual with a duality it is hard
to prove, since calculations can only be done at weak coupling
in the gauge theory, $g^2N\ll1$, while---presently at
least---calculations
on the string theory side can only be done in the classical
supergravity limit where the radius of curvature $R\gg\sqrt{\alpha'}$;
which means $g^2\ll1$ while $g^2N\gg1$.
Some quantities, however, are protected
against renormalization in $g^2N$, and the value calculated in the
gauge theory can be compared directly to the value extracted from the
supergravity approximation to string theory.

For us the relevant correlation functions involve 16 dilatinos
$\Lambda$ on the
supergravity side that correspond to a certain composite operator in
the gauge theory.
These correlation functions receive contributions from D-instantons in
the string theory whose coupling
dependence singles them out as instanton contributions in the gauge
theory. The correspondence requires that the $k$ instanton contribution to the
correlator, in the infra-red and at leading order in $\tfrac1N$,
should be, schematically, \cite{BG,Bianchi:1998xk}
\SP{
&\Vev{\Lambda(x_1)\cdots \Lambda(x_{16})}\thicksim
\sqrt N\,g^{-24}\,q^k\,k^{25/2}\,\sum_{d|k}d^{-2}\,\label{16fc}\\
&\qquad\qquad\times\int \frac{d^4X d\rho}{\rho^5}\prod_{i=1}^{16}
{\cal F}(x_i-X,\rho)\ .
}
Here, $\{X_n,\rho\}$ parameterizes a point in $AdS_5$ and the details of
the expression for the integrand may be found in
\cite{MO3}. What is remarkable about \eqref{16fc} is that the $k$
dependence is only through the numerical pre-factor
$q^kk^{25/2}\sum_{d|k}d^{-2}$. This looks like a disaster because
there seems little chance that the integral over $k$ instantons, with
its intrinsic complexity, would reduce to something that is simply a number
times a one instanton contribution.

The $k$-instanton contribution to the correlators involves inserting
into the measure, \eqref{meas} with $\N=4$ along with the 4-fermion
coupling \eqref{4fc}, the 16 composite operators.
This contribution turns out to be
calculable at leading order in $1/N$
in a way that we summarize below \cite{MO3}:

(i) For $N\geq2k$, and
so certainly at large $N$, the ADHM constraints can be solved by a
simple change of variables: the biggest impediment to progress with
the ADHM construction proves to be entirely benign. The idea involves
changing variables from the $w_\aD$ to quadratic gauge invariant
variables
\EQ{
W^\aD_{\ \bD}=\bar w^\aD w_\bD\ .
}
The ADHM constraints are then linear in $W^\aD_{\ \bD}$, as is
apparent from \eqref{ADHM}, and the the $\delta$-function constraints
in \eqref{meas} may trivially be solved.

(ii) The 4-fermion term in the instanton action \eqref{4fc} can be
bilinearized by introducing a 6-vector of $U(k)$-adjoint variables
$\chi_a$. The Grassmann collective coordinates can then
integrated-out. The $\chi_a$ variables are precisely those that arise
naturally in the D-instanton/D3-brane system described previously.

(iii) The remaining expression is then amenable to a
saddle-point approximation at large $N$. The saddle-point solution has
a very simple interpretation. Each of the $k$ instantons are embedded
in mutually commuting $SU(2)$ subgroups of the gauge group, as one
might have expected on statistical grounds alone. Furthermore, and
less intuitive, is that they have
the same size $\rho$ and sit at the same point $X_n$ in
spacetime; so at the saddle point
\EQ{
\bar w^\aD w_\bD=\rho^2\delta^\aD_{\ \bD}1_{\sst[k]\times[k]}\ ,\qquad
a'_n=-X_n1_{\sst[k]\times[k]}\ .
}
Furthermore, the auxiliary variables $\chi_a$ have the saddle-point value
\EQ{
\chi_a=\rho^{-1}\hat\Omega_a1_{\sst[k]\times[k]}\ ,
}
where $\hat\Omega_a$ is a unit 6-vector. So the saddle point is
parameterized by a point in $AdS_5\times S^5$! Amazingly, instantons in
the gauge theory act as a probe that feel the ten-dimensional geometry
of the dual theory.
Notice that the $S^5$ part of the geometry arises from the auxiliary
variables $\chi_a$.

(iii) The integral of the fluctuations
around the saddle-point solution assembles
into something that is known: precisely the partition function of
$\N=1$ 10D $SU(k)$ Yang-Mills dimensionally reduced to 0 dimensions,
where the 10D gauge field is formed from the traceless parts of $a'_n$
and $\chi_a$. This is known to be proportional to $\sum_{d|k}d^{-2}$
\cite{Moore:2000et,KNS}.

Putting all of this together
immediately solves the puzzle alluded to above: any correlation
function will look one instanton-like up to an overall $k$ dependent
factor. In addition, one can show that the $k$-dependence and overall
factor of $\sqrt N$ are exactly reproduced. Instantons consequently
provide one of the
most convincing pieces of evidence in favour of the AdS/CFT correspondence.

We can also couch our result in terms of the partition
function of the D-instanton/D3-brane system:
\SP{
&\Z_{k,N}\underset{N\to\infty}=
2^{3-2k}\pi^{6k-25/2}\\ & \sqrt
Nk^{3/2}\sum_{d|k}d^{-2}\int\frac{d^4X\,d\rho\,d^5\hat\Omega}{\rho^5}\cdot
d^8\xi\,d^8\bar\eta\ ,
\label{lnr}
}
where $X$ and $\rho$ are the overall position and scale size,
respectively, while
$\xi$ and $\bar\eta$ are the 8 SUSY and superconformal fermion
zero modes, respectively. It is possible to generalize these kinds of
calculations to other AdS/CFT duals \cite{others}.

\subsection{Instanton effects in D3-branes}\label{sec:S6}

The collective excitations of $N$ D3-branes are described at low
energies by
an $\N=4$ SUSY gauge theory with gauge group $U(N)$. However, the
minimal $\N=4$ Lagrangian is only valid at low energy and there is an infinite
tower of the higher derivative interactions
that come with powers of $\alpha'$, the string length scale.
Some of these, {\it but not all} are encoded in the
Born-Infeld Lagrangian. In
\cite{GG}, it was argued that in the case of a single D3-brane,
instantons contribute to certain terms of
order $\alpha^{\prime4}$, including
one of the
form $(\partial F)^4$, where $F$ is the abelian field strength.
Furthermore, the $SL(2,{\mathbb Z})$ modular symmetry of the Type IIB 
string theory, which is realized as electro-magnetic duality in the
D3-brane theory, fixes
the instanton contributions exactly. In fact the coupling to this term
in the effective action involves the logarithm of the Dedekind eta function:
\EQ{
\ln|\eta(\tau)^4|=-\tfrac\pi3{{\rm Im}\tau}-2\sum_{k=1}^\infty(q^k+\bar
q^k)\Big[\sum_{d|k}d^{-1}\Big]\ .
}
Here, the first term is a tree-level contribution, while the other
terms come from $k$ instantons and $k$ anti-instantons, respectively.

We can relate the $k$-instanton terms in the effective action of the D3-brane
predicted by Green and Gutperle \cite{GG} for the case
$N=1$ (without FI and VEV terms) to the $k$-instanton partition function
modded out by the
integral over the overall $k$-instanton position
in ${\mathbb
R}^4$ and its superpartners (the 8 supersymmetric
fermion zero modes)
\EQ{
\widehat\Z_{k,1}(\zeta,\alpha')=\sum_{d|k}d^{-1}\ .
\label{ggr}
}
Here, the FI coupling $\zeta$, absent at the start, arises as a source.

What is interesting about the string result \eqref{ggr}
is that in order to have a
non-trivial contribution when $\zeta=0$, it is absolutely essential to have the
$\alpha'$ corrections in the D-instanton/D3-brane system. Another way of
seeing this is that superconformal invariance must be broken. It turns
out that when $\zeta\neq0$
we can legitimately set $\alpha'=0$ to yield:
\EQ{
\widehat\Z_{k,1}(\zeta,0)=\sum_{d|k}d^{-1}\ ,
}
which is then a statement about the integral over the resolved {\em
centered\/} moduli
space $\widehat\ms_{k,1}^{(\zeta)}$, where $\ms_{k,N}^{(\zeta)}={\mathbb
R}^4\times\ms_{k,N}^{(\zeta)}$.
Note that the integral \eqref{ggr} does not actually
depend on the $\alpha'$ coupling; a fact that we shall return to in
\S\ref{sec:S7}.

The whole story of D-instanton corrections to the D3-brane
effective action generalizes to the non-abelian case of $N$ D3-branes
\cite{DHK}. In this
case, it is necessary that the D3-branes are separated by adding VEVs
$\langle\varphi\rangle$ so that
theory is in a Coulomb phase. In this case, there is generalization of
\eqref{ggr}:
\EQ{
\widehat\Z_{k,N}(\zeta,\alpha',\langle\varphi\rangle)=N\sum_{d|k}d^{-1}\ .
\label{ggn}
}
It is possible to prove this using Morse theory
arguments. First of all, one can set $\alpha'$ to zero in \eqref{ggn}.
If the VEVs vanished, the latter
quantity is Gauss-Bonnet-Chern integral on
$\widehat\ms_{k,N}^{(\zeta)}$ (see the next section).
Turning on the VEV has the effect of introducing a Morse
function on $\widehat\ms_{k,N}^{(\zeta)}$ and using standard arguments
the integral $\widehat\Z_{k,N}(\zeta,0,\langle\varphi\rangle)$ localizes onto the
critical point set. These are the submanifold of $\ms_{k,N}^{(\zeta)}$
where
\EQ{
w_\aD\chi_a+\langle\varphi_a\rangle w_\aD=0\ .
}
The VEV $\langle\varphi_a\rangle$ is a 6-vector of diagonal $N\times N$ matrices. The
critical points correspond to associating each instanton with a
particular D3-brane; in other words a partition
$k=\{k_1,\ldots,k_n\}$, where $k_u\geq0$. For each partition the
critical point set is a product of abelian, $N=1$, instanton moduli
spaces:
\EQ{
\ms_{k_1,1}\times\cdots\times\ms_{k_n,1}\ .
}
Localizing the integral on
the critical point sets gives us a relation of the form
\EQ{
\Z_{k,N}(\zeta,0,\langle\varphi\rangle)=\sum_{\{k_j\}}
\Z_{k,1}(\zeta,0)\times\cdots\times\Z_{k,1}(\zeta,0)\ .
}
Notice that we have not separated out the center-of-mass integrals
yet. Each factor $\Z_{k,1}(\zeta,0)$ leaves 8 unsaturated
Grassmann integrals; hence, most of the partitions in the sum have
more than 8 unsaturated Grassmann integrals and will not contribute
to $\widehat\Z_{k,N}(\zeta,0,\langle\varphi\rangle)$. Only the partitions where all $k$ of
the instantons live on the same D3-brane will survive, and there are
$N$ of these; so
\EQ{
\widehat\Z_{k,N}(\zeta,0,\langle\varphi\rangle)=N\widehat
\Z_{k,1}(\zeta,0)=N\sum_{d|k}d^{-1}\ .
}
What is striking about this result is that for $N>1$ it
holds only for nonvanishing VEVs, {\it i.e.\/}~in the Coulomb phase, but 
nevertheless the right-hand side of \eqref{ggn} is independent of
$\langle\varphi\rangle$. 

\subsection{The Euler Characteristic of $\widehat\ms_{k,N}$}\label{sec:S7}

This section investigates the relation between various physical
quantities extracted from the brane system and the
Euler characteristic of instanton moduli space \cite{DHK}. The 
Euler characteristic $\chi$ of resolved (centered) moduli space
$\widehat\ms_{k,N}^{(\zeta)}$ was deduced 
from the Morse theory analysis of Nakajima
\cite{Nak} (see also \cite{Aharony:1998an}):
\EQ{
1+\sum_{k=1}^\infty\chi(\widehat\ms_{k,N}^{(\zeta)})q^k=
\frac1{\prod_{j=1}^\infty(1-q^j)^N}\ .
\label{hom}
}
Notice that the generating function above is, up to a factor of
$q^{-N/24}$,  $\eta^{-N}$, where $\eta$ is the Dedekind
eta-function and, as such, has interesting modular properties. In fact
there is a very intriguing more general relation between the
generating functions of the Euler characteristics for instantons in
gauge theories defined on different 4 manifolds, and the characters of
2-dimensional conformal field theories \cite{VW}.
The question is how this relates to the
the resulting partition function of the D-instanton/D3-brane system,
with the centre-of-mass and the 8 SUSY zero modes factored off, denoted
\EQ{
\widehat\Z_{k,N}(\zeta,\alpha',\langle\varphi\rangle)\ .
}
where the dependence on  FI couplings, VEVs and $\alpha'$ is
indicated.

Conventional wisdom suggests that the
quantity $\widehat\Z_{k,N}$ should yield the Euler characteristic
of $\widehat\ms_{k,N}$ as we shall now explain. The
point is that the measure over the instanton moduli space in an $\N=4$
gauge theory has the 4-fermion term \eqref{4fc}. The 4-index tensor
that appears is nothing but the Riemann tensor of $\widehat\ms_{k,N}$ and
saturating the fermion integrals from the action
brings down powers of the Riemann tensor
contracted in such a way that the resulting integral over $\widehat\ms_{k,N}$
is precisely the Gauss-Bonnet-Chern (GBC) integral. On a compact manifold this
would give the Euler characteristic. However, our manifold is non-compact
and in this case the GBC integral only gives the bulk contribution to the
Euler characteristic and there can be additional boundary contributions arising
from either the boundary at infinity or from the orbifold singularities.

One immediate problem with our integral is that
due to superconformal invariance the integral
over the overall scale size of the configuration is divergent. In
addition the integrals over the 8 superconformal zero modes are not
saturated by any fermionic insertions. This is evident from our large-$N$
expression for
$\Z_{k,N}$ in \eqref{lnr}. The way to deal with this
is to break superconformal invariance by blowing up the singularities
of $\widehat\ms_{k,N}$ to give the smooth, but still non-compact, space
$\widehat\ms_{k,N}^{(\zeta)}$. This is achieved by
including the FI couplings $\zeta$.
The integral over the scale size is now rendered convergent and the
8 superconformal zero modes are lifted. Factoring off the integrals over
the overall position and 8 SUSY zero modes,
gives the GBC integral in the large $N$ limit:
\EQ{
\widehat\Z_{k,N}(\zeta,0,0)\underset{N\to\infty}=
2^{3-2k}\pi^{6k-13/2}\sqrt
Nk^{3/2}\sum_{d|k}d^{-2}\ .
}
We can also calculate the $k=1$ case with arbitrary $N$ exactly \cite{DHK}:
\EQ{
\widehat\Z_{1,N}(\zeta,0,0)=\frac{2^{1-2N}(2N)!}{N!(N-1)!}\
.
\label{onen}
}
These expressions are not integers and there must be contributions
coming form the boundary at infinity. In particular, on the basis of
the formula \eqref{hom}, we expect $\chi(\widehat\ms_{1,N}^{(\zeta)})=N$.
For $k=1$ and $N=2$ we can calculate the surface contribution and show how the
\eqref{hom} calculation is consistent with \eqref{onen}. In this case
$\widehat\Z_{1,2}(\zeta,0,0)=\tfrac32$, while $\chi=2$. Recall
that the singular space
$\widehat\ms_{1,2}={\mathbb R}^4/{\mathbb Z}_2$ while the blow-up
$\widehat\ms_{1,2}^{(\zeta)}$ is the Eguchi-Hanson
manifold. In  \cite{Eguchi:1980jx},
the authors calculate the Euler
characteristic of this space
and show that it receives a bulk contribution of $\tfrac32$, matching
\eqref{onen},
and---importantly for us---a boundary contribution of $\tfrac12$.

For the case $k=1$ at least (and arbitrary $N$)
one can alter the asymptotic behaviour of
the GBC integral so that there is no contribution from infinity and
the bulk integral yields directly the Euler characteristic $\chi=N$. This
modification involves moving onto the Coulomb branch of the $\N=4$
theory by introducing VEVs as in \eqref{vev}. We then find by explicit
calculation \cite{DHK}
\EQ{
\widehat\Z_{1,N}(\zeta,0,\langle\varphi\rangle)=N\ .
\label{intv}
}
Interestingly, moving out onto the Coulomb branch can
be interpreted as introducing a
potential on the moduli space. This potential has precisely
$N$ critical points whose number saturates the value of the integral as
one might have expected from standard
Morse theory arguments. In fact, in a similar way 
for general $k$, there is a more
refined potential that arises from giving all the fields {\it twisted
masses\/} \cite{DHK}, which lifts all the flat directions and has a number of
isolated critical points which yields directly the result of Nakajima 
\cite{Nak}.
To complete the story of $k=1$ we can also calculate explicitly
\SP{
&\widehat\Z_{1,N}(0,0,\langle\varphi\rangle)=N-\frac{2^{1-2N}(2N)!}{N!(N-1)!}\\
&\equiv \widehat\Z_{1,N}(\zeta,0,\langle\varphi\rangle)-\widehat\Z_{1,N}(\zeta,0,0)\ .
}
It remains to be seen whether similar relations hold for
arbitrary $k$;  however, we feel that it should be possible to relate
$\widehat\Z_{k,N}(\zeta,0,\langle\varphi\rangle)$ and $\widehat\Z_{k,N}(0,0,\langle\varphi\rangle)$
by investigating the behaviour of the integrals near the
singularities.
The quantity $\widehat\Z_{k,N}(0,0,\langle\varphi\rangle)$ is particularly interesting
because it appears in the mismatch between the microscopic couplings
and the effective couplings in Seiberg-Witten theory of the mass deformed
$\N=4$ theory \cite{Dorey:1997ez}.

Up till now, we have not considered the effect of the stringy
couplings. Actually, turning on the $\alpha'$
couplings has no effect; in other words,
\EQ{
\widehat\Z_{k,N}(\zeta,\alpha',\langle\varphi\rangle)=\widehat\Z_{k,N}(\zeta,0,\langle\varphi\rangle)\ .
}
as long as either $\zeta$ or $\langle\varphi\rangle$
are non-vanishing so that the right-hand side is well defined.
We have already seen in \eqref{ggn} this is implied by a
generalization of the Green-Gutperle analysis to $N>1$.
There is another way to motivate this result from completely different
considerations involving ideas of Cohomological Field Theory
\cite{Moore:2000et}. In a very small nut-shell,
the integral $\widehat\Z_{k,N}(\zeta,\alpha',\langle\varphi\rangle)$ has
a nilpotent BRST-type symmetry derived from its
supersymmetry. The $\alpha'$ couplings can be shown to be $Q$-exact
and hence can be set to 0, as long as the resulting integral is
well defined.

There is clearly a lot of interesting relations between
the Euler characteristic of
instanton space and various physical quantities in the gauge/string
theory that remain to be uncovered.

We would like to thank Stefan Vandoren for conversations and
collaboration and  
our erstwhile collaborator Michael Mattis. This work was supported by
a TMR network, EC contract number FMRX-CT96-0012.

\end{document}